# Modeling Dynamics of Complex System with Solutions of the Generalized Lotka-Volterra Equations

Lev A Maslov

UNC, Aims College, Greeley, CO

lev.maslov@cccs.edu

**Abstract**

A system of nonlinear ordinary differential equations with forcing function is developed to model evolution processes in complex systems:

$$\left[\begin{array}{l} \dfrac{dR}{dt} = -a_1 \cdot R \cdot C + a_2 \cdot R \cdot P \\ \dfrac{dC}{dt} = -a_3 \cdot C + a_4 \cdot C \cdot R + F \\ \dfrac{dP}{dt} = a_5 \cdot P - a_6 \cdot P \cdot R \end{array}\right]$$

In this system *R, C,* and *P* are the "resource", "consumption", and "production" functions correspondingly. *F* is the forcing function. Two cases *F≡0*, and *F=A+Bsin(ωt)* are considered. It is shown that if *F≡0*, there exist an unstable periodic solution for a certain set of coefficients $a_i$. In the case of periodic forcing, system has an attractor solution for a certain set of coefficients $a_i$, *i=1..6*.

The system developed may have a wide range of applications: in biological and social sciences, in economics, in ecology, and, as well, in modeling climate. A correspondence between the theoretical (numerical) solution and Late Pleistocene climate data variations is analyzed on a qualitative level.

**Introduction**

In an extensive review M. Crucifix (2012) discussed a number of mathematical models of climate dynamics: (Saltzman and Maasch, 1988; Paillard and Parrenin, 2004) and some others. In addition, we would like to mention here a mathematical model of climate developed by V. Sergin (Sergin, 1979). A system of differential equations representing this model was written and solved numerically. It was shown that there exist an auto-oscillation in a system with periods of 20,000-80,000 years. Another mathematical model of the interaction between continental ice, ocean and atmosphere in a form of a system of ODE was developed by B. Kagan and coauthors (Kagan, Maslova, Sept, 1993). Given realistic thermodynamic parameters, the authors found an auto-oscillation within this system, with a period of about 100,000 years.

The most important feature of all the mathematical models mentioned (and not mentioned) above is that they are made **ad hoc**, to model some specific **climate** situations. The goal of the current work is to present more general than simple Lotka-Volterra equations, mathematical construction capable to model dynamics of complex systems.

**The basic equations**

$$\begin{bmatrix} \dfrac{dR}{dt} = -a_1 \cdot R \cdot C + a_2 \cdot R \cdot P \\ \dfrac{dC}{dt} = -a_3 \cdot C + a_4 \cdot C \cdot R + F \\ \dfrac{dP}{dt} = a_5 \cdot P - a_6 \cdot P \cdot R \end{bmatrix} \qquad (1)$$

with initial conditions $R(0) = R_0, P(0) = P_0, C(0) = C_0$, and with constant coefficients $a_1, a_2, a_3, a_4, a_5, a_6 > 0$. $F$ is the forcing function, $F = A + B \cdot \sin(\omega \cdot t)$. This system was presented first in (Maslov, 2014).

*A, B, ω* are constant and positive, *t* - time. *R* is the "resource" function of a system. In the first equation of (1) functions *P* and *C* form the "budget" *R* of a system. Function *P* increases rate of change of *R*, while function *C* decreases rate of change of *R*. The second equation of (1) describes evolution of function *C*. The first term in the RHS of this equation represents the rate of decay of this function, while the second term represents the rate (positive) of change of *C* due to interaction with *R*. The third equation in (1) describes evolution of *P.* The first term in the RHS of this equation represents the rate of growth of this function, while the second term represents the rate (negative) of change of *P* due to interaction with *R*.

Contrary to Lotka-Volterra equations, functions *P* and *C* do not interact directly. They interact through the resource function *R* which they create in the process evolution of a system. The classical Lotka-Volterra system of equations is a particular case of system (1) for $a_1 = a_2 = 0$.

**Interpretation of equations (1)**

The resource function *R* can be interpreted in many different ways. For instance, it can be interpreted as amount of energy (or fraction of it) stored in a system. In this case, *dC/dt*, and *dP/dt* are rates of transformation energy of a system into different forms due to the internal processes. Functions *P(t)* and *C(t)* can be interpreted as, for instance, "entropy", and "temperature" of a system.

**The balanced and unbalanced solutions of (1) for F≡0**

The first set of critical points of the system (1) is $R = P = C \equiv 0$. The second set of critical points we obtain from the system

$$\begin{bmatrix} a_1 \cdot C - a_2 \cdot P = 0 \\ -a_3 \cdot (1 - \dfrac{a_4}{a_3} \cdot R) = 0 \\ +a_5 \cdot (1 - \dfrac{a_6}{a_5} \cdot R) = 0 \end{bmatrix} \qquad (2)$$

The system (2) is compatible if $\dfrac{a_4}{a_3} = \dfrac{a_6}{a_5}$.

In this case, the system (1) has a periodical solution, Figure 1.

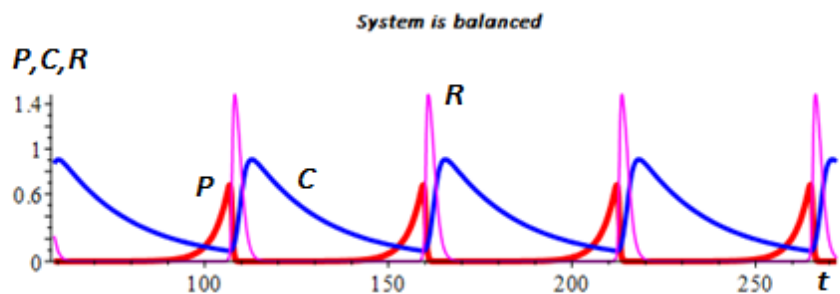

Figure 1. Solution of (1) for $a_4 / a_3 = a_6 / a_5$.

This process is, however, unstable in the sense that a periodic solution does not exist for any $\varepsilon \neq 0$ in

$$\frac{a_4}{a_3} = \frac{a_6}{a_5} + \varepsilon$$

Figure 2 shows three solution of (1) for a balanced system, (*a*); for $a_3$ which is greater than $a_3$ of a balanced system (*b*); and for $a_3$ which is less than $a_3$ of a balanced system (*c*).

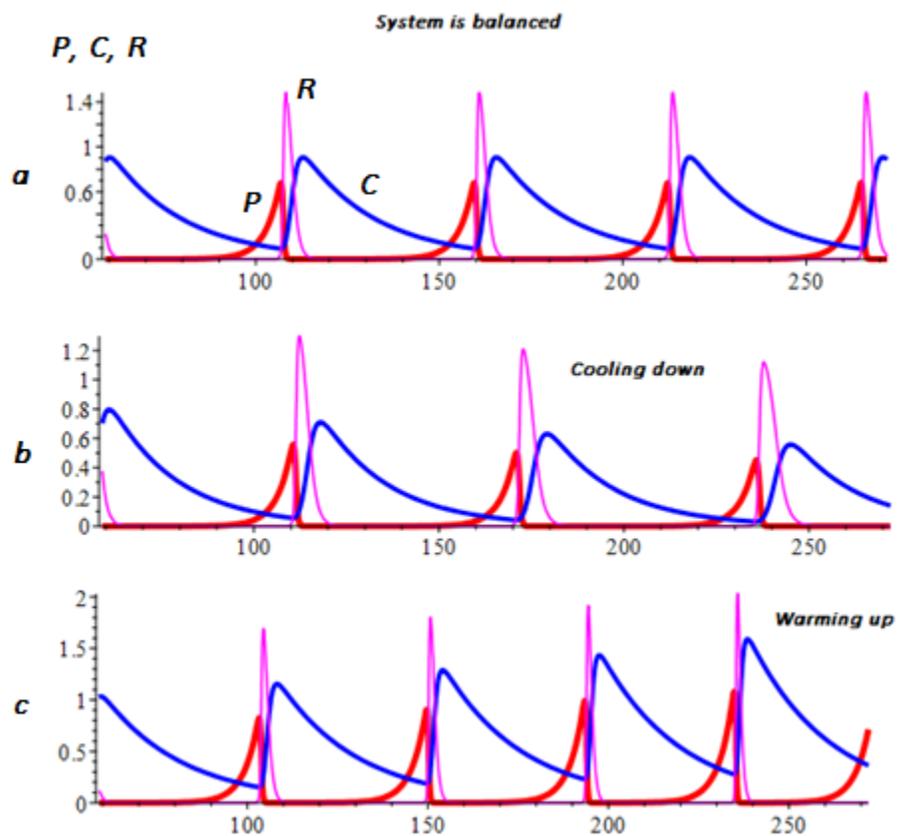

Figure 2. Solution of (1) for a balanced system (*a*); $a_3$> $a_3$ of a balanced system (*b*), and for $a_3$< $a_3$ of a balanced system (*c*).

Exponential growth of *P(t)* (red) governed by coefficient $a_5$ and exponential decay of *C(t)* (blue) governed by coefficient $a_3$ are observed on these graphs. Coefficients $a_3$ and $a_5$ can be interpreted as regulators of rate of dissipation and as rate of accumulation of energy in a system.

**Interpretation of theoretical and observed data**

Figure 3 shows variation of temperature *T* in four Late Pleistocene climate cycles (Petit, et al., 1999, 2001), and graphs of natural logarithm of *T*. The straight line approximation of *ln(T)* suggests the exponential decay of temperature.

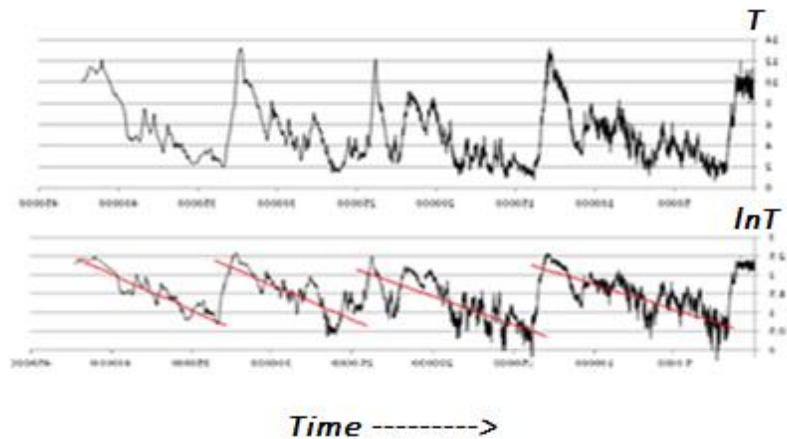

Figure 3. Temperature *T* data, and *ln(T)* with approximation

*ln(T) = -k t + p* within cycles.

From the other side, the straight line approximation of natural logarithm of dust concentration, Figure 4, suggests the exponential growth of dust concentration in each of four Late Pleistocene climate cycles, Figure 4.

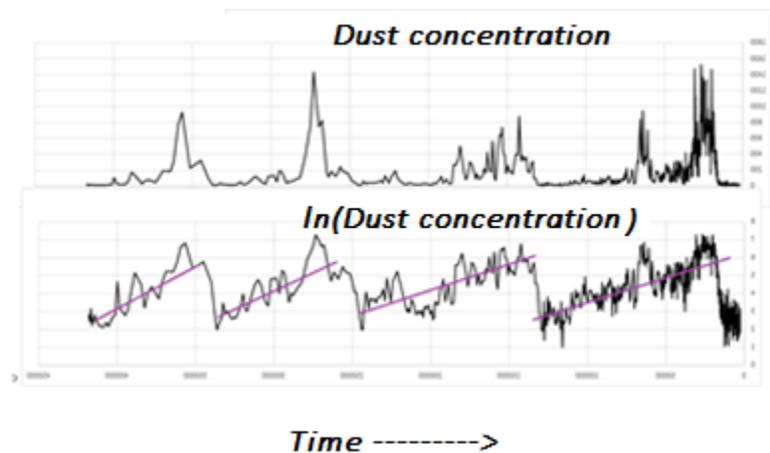

Figure 4. *Dust concentration* data (Petit, et al., 1999, 2001), and approximations

of *ln (Dust concentration) = s t + d* within cycles.

A comparison of temperature *T* and *Dust concentration* data on the same time scale is shown in Figure 5. It is seen from this Figure that peaks of dust concentration precede peaks of temperature *T* over a relatively short time interval *δt*. Exponential increase of dust concentration, and exponential decrease of temperature *T*, as demonstrated on Figures 3 and 4, are observed in all of four cycles.

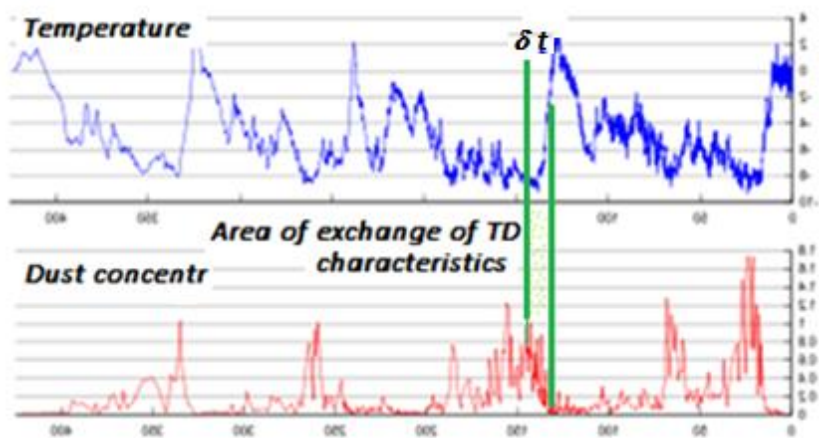

Figure 5. *Temperature* variation and *Dust concentration* data in Late Pleistocene

(Petit, et al., 1999, 2001)

The similar dynamics pattern is observed on graphs of *P(t)* and *C(t)* functions, Figure 6. Peaks of *P(t)* precede peaks of *C(t)*. Interaction of *P*(t) and *R*(t), causes decrease of *P*(t) and sharp increase of *R*(t), left boundary. In turn, interaction of *R*(t) and *C*(t) causes decrease of *R*(t) and sharp increase of *C*(t), right boundary of the interval.

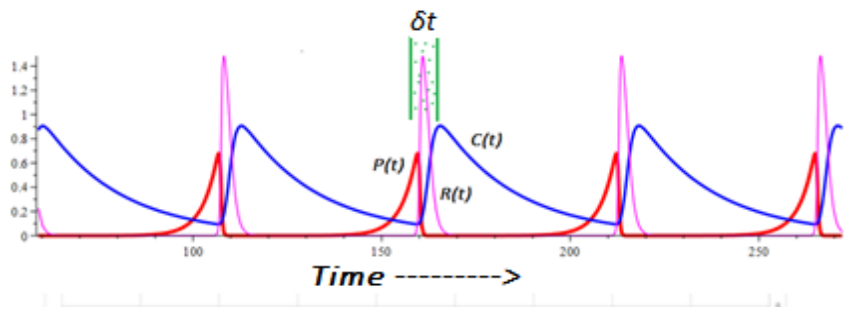

Figure 6. The time interval *δt* between moments of interaction of *P*(t) and *R*(t), left boundary, and interaction of *R*(t) and *C*(t) functions, the right boundary of the interval.

The similarity between theoretical solution of (1) dynamics and the Late Pleistocene climate data dynamics requires explanation and thermodynamic interpretation of functions *P(t), R(t)* and *C(t)*.

**Solutions with forcing function *F=A + Bsin(ωt)***

Behavior of a system (1) with forcing function *F* requires special investigation. In this paragraph we give a few illustrations characterizing properties of a system on a qualitative level. Figure 7 shows the gradual increase of *C(t)* and *P(t)* of the originally balanced system with forcing function *F=A + Bsin(ωt)* added; *A > 0, B > 0*.

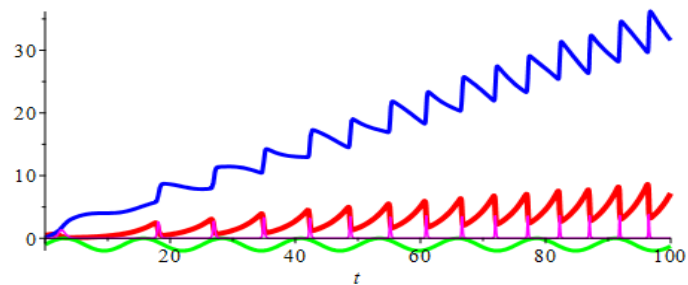

Figure 7. Evolution of a balanced system $a_4/a_3 = a_6/a_5$ with added forcing function *F=A + Bsin(ωt), A>0, B>0*.

However, the steady evolution of a system is possible if to take coefficient $a_3$ balancing input of *A* (solar irradiation, for instance).

Figure 8 shows *R, C* and *P* quasiperiodic oscillations for $|a_3 \cdot C| \approx A$, and $a_4/a_3 \neq a_6/a_5$.

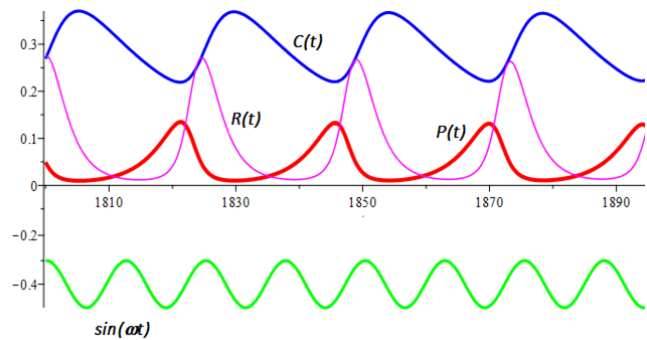

Figure 8. The same system as on Figure 7, but with coefficient $a_3$ such that $|a_3 \cdot C| \approx A$

If to take coefficient $a_3$ much greater than that of a balanced system, the fast decrease of a system auto oscillation can be observed, followed then by oscillations of the forcing function.

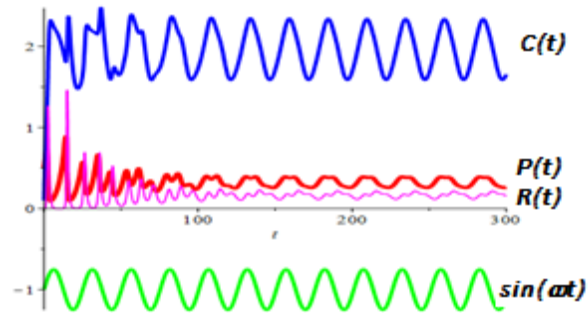

Figure 9. Dynamics of a system in case the coefficient $a_3 >>$ ( $a_3$ of a balanced system),

and the forcing function $F=A + Bsin(\omega t)$, $A>0$, $B>0$.

It is observed here the fast decrease of Eigen oscillations of a system, followed by oscillations synchronous with that ones in the forcing function.

If $B=0$ in $F=A + Bsin(\omega t)$, then

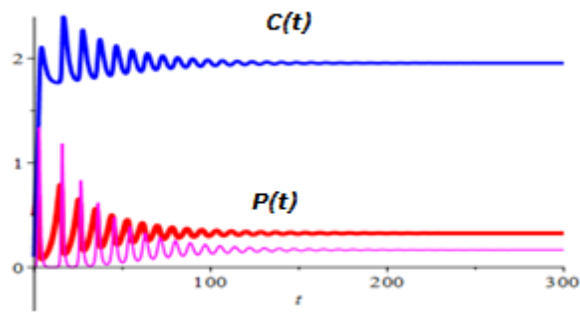

Figure 10. Dynamics of a system if $A>0$ and $B=0$ in the forcing function $F=A + Bsin(\omega t)$,

and $a_3 >>$ ($a_3$ of a balanced system).

For a certain set of parameters (coefficients of a system), the interaction between Eigen oscillation of a system and forcing function oscillations can be more complex, see Figure 11.

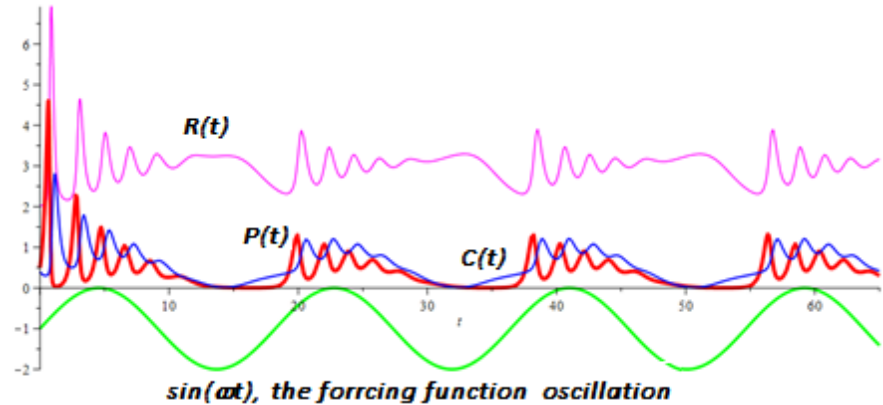

Figure 11. Dynamics of a system if *A>0, B>0,* and  $a_3$ >>  ($a_3$ of a balanced system).

Can we treat this, and similar solutions, of the system (1) as attractors? Numerical experimenting showed that there exists a region of initial conditions (basin of attraction) such that all solutions with initial conditions from this region converge asymptotically to a solution, Figure 11, attractor of a system, for a certain set of coefficients. Thus, the system (model) developed is capable of synchronizing its oscillations with oscillations of forcing function.

**Discussion and conclusions**

A system of nonlinear ordinary differential equations (1), generalization of Lotka-Volterra equations, is developed to model evolution processes in complex structures. This system may have a wide range of applications: in economics, in biological and social sciences, in ecology, and, as well, in modeling climate dynamics.

It is shown that theoretical solution of (1) models the Late Pleistocene climate data dynamics on qualitative level very well. This result requires explanation and thermodynamic interpretation of *P*(t), *R*(t) and  *C*(t) function. A special mathematical research has to be undertaken to study properties of a system (1).

The content of this paper was presented at Prof. André Berger seminar 10/24/2017, University Catholic Louvain, Belgium.

101010